\documentclass[twocolumn,twoside,slac_two]{revtex4}
\usepackage{graphicx}
\usepackage{fancyhdr}
\newcommand{\sun}{\ensuremath{\odot}}
\def\ltsima{$\; \buildrel < \over \sim \;$}
\def\simlt{\lower.5ex\hbox{\ltsima}} 
\def\gtsima{$\; \buildrel > \over \sim \;$}
\def\simgt{\lower.5ex\hbox{\gtsima}} 

\def\phflux{ph cm$^{-2}$ s$^{-1}$}
\def\deg{\hbox{$^\circ$}}

\def\p0{$\pi^{\rm 0}$}

\def\Fermi{\textit{Fermi}}
\def\Swift{\textit{Swift}}

\begin{document}

\title{\Fermi\ Discovers a New Population of Gamma-ray Novae}

\author{C.C. Cheung}
\affiliation{Space Science Division, Naval Research Laboratory, Washington, DC 
20375, USA\\
on behalf of the \textit{Fermi}-LAT Collaboration}

\begin{abstract} Novae had not been widely considered as high-energy ($>100$ 
MeV) gamma-ray sources before the launch of the \Fermi\ Large Area Telescope 
(LAT). In March 2010, the LAT made the first gamma-ray detection of a nova in 
the symbiotic binary V407 Cygni. The LAT observations uniquely probed the 
high-energy particle acceleration mechanism in the environs of the V407 Cyg 
binary system consisting of a white dwarf and its red giant companion. 
Subsequently in June 2012, two more novae were detected with the LAT, Nova Sco 
2012 and Nova Mon 2012, thus heralding novae as a new gamma-ray source class. 
For Nova Mon 2012, the gamma-ray transient source was discovered first, followed 
by the optical confirmation of the nova, showcasing how novae can be found with 
the LAT independently from traditional optical searches. We discuss the LAT 
detected gamma-ray novae together with observational limits on other optical 
novae over the first four years of the \Fermi\ mission and reconsider the 
possible high-energy gamma-ray production mechanisms in novae in light of the 
new detections. \end{abstract}

\maketitle

\thispagestyle{fancy}

\section{\Fermi\ Discovery of Nova V407 Cyg -- the First Surprise}

The \Fermi\ Large Area Telescope (LAT) maps the entire sky every $\sim 3$ hrs 
with unprecedented sensitivity and localization capabilities in the high-energy 
(HE; $>100$ MeV) gamma-ray energy range. The data are sensitive to and 
continuously searched for flaring/transient sources by a dedicated group of 
flare advocates \cite{cip12} in the \Fermi-LAT team via an Automated Science 
Processing (ASP) \cite{atw09} pipeline. This surveying capability leads 
naturally to a search of bright variable gamma-ray sources in the Galactic 
plane, and has been pursued since early in the mission \cite[e.g.,][]{hay09}.

As part of this flare advocate ASP program, in 2010, a new gamma-ray source was 
detected by the LAT in the Cygnus region of our Galaxy. Initially dubbed Fermi 
J2102+4542, the source was bright, being detected over two consecutive days 
(March 13th and 14th) with $>100$ MeV fluxes $\simgt 10^{-6}$ ph cm$^{-2}$ 
s$^{-1}$. Within the LAT localization error was the symbiotic star V407 Cyg, 
whose optical outburst detection on March 10 \cite{nis10} triggered \Swift/XRT 
observations on March 13th and 15th. X-ray emission from V407 Cyg was detected 
in the \Swift\ observations for the first time, and as importantly, no other 
X-ray source was found within the LAT error circle. These observations motivated 
the proposed association of the gamma-ray transient with V407 Cyg \cite{che10}, 
but HE gamma rays from novae were not widely anticipated, so this initially met 
with some skepticism. However, subsequent analysis found the initial (fainter) 
LAT detection was indeed on March 10, the day of the optical nova discovery. 
This temporal coincidence together with the LAT localization, provided 
convincing evidence that the gamma-ray emission originated from the nova 
\cite{abd10}.

The basic picture in the case of V407 Cyg was that high-energy particles could 
be accelerated in the shock between the nova ejecta and the wind of the red 
giant (RG) companion as had been proposed in the symbiotic recurrent novae RS 
Oph by \citet{tat07}; see also \cite{her12}. The gamma-ray production could be 
due to either \p0\ decay produced in interactions of relativistic protons with 
the material of the RG, or relativistic electrons that interact with the photons 
and wind of the RG by inverse Compton and bremsstrahlung emission, respectively 
\cite{abd10}; see also \cite{mar13}. The scenarios outlined were however, 
particular to the symbiotic nature of this nova, thus provided a natural 
explanation for the non-detection of bright gamma rays from the more commonly 
observed class of classical novae (CN; Table~1) up to that time.

\begin{table}[t]
\begin{center}
\caption{\textbf{: White Dwarfs in Close Binary Systems}}
\begin{tabular}{|c|c|c|}
\hline \textbf{} & \textbf{Classical} & \textbf{Symbiotic} \\
 &  \textbf{Novae} & \textbf{Recurrent Novae} \\
\hline
System  & compact CV-like & symbiotic-like \\
(Binary) & (WD + Main Sequence) & (Massive WD + RG) \\
\hline
$a$ & $10^{11}$ cm $\sim R_{\sun}$ & $\sim$100's $R_{\sun}$ \\
$P_{\rm rec}$ & $\simgt 10^{4}$ yrs & $< 100$ yrs \\
$P_{\rm orb}$ & $\sim$ hr - day & $\sim$years \\
Rate & $\sim 35$/yr in Galaxy & $\sim 10$ known \\
\hline
\end{tabular}\label{table1}
Summary adapted from \cite{her11} and modified for this talk.
Cataclysmic Variable (CV), binary separation ($a$), recurrence 
period ($P_{\rm rec}$), orbital period ($P_{\rm orb}$).
\end{center}
\end{table}

\begin{figure*}[t]
\centering
\includegraphics[width=88mm]{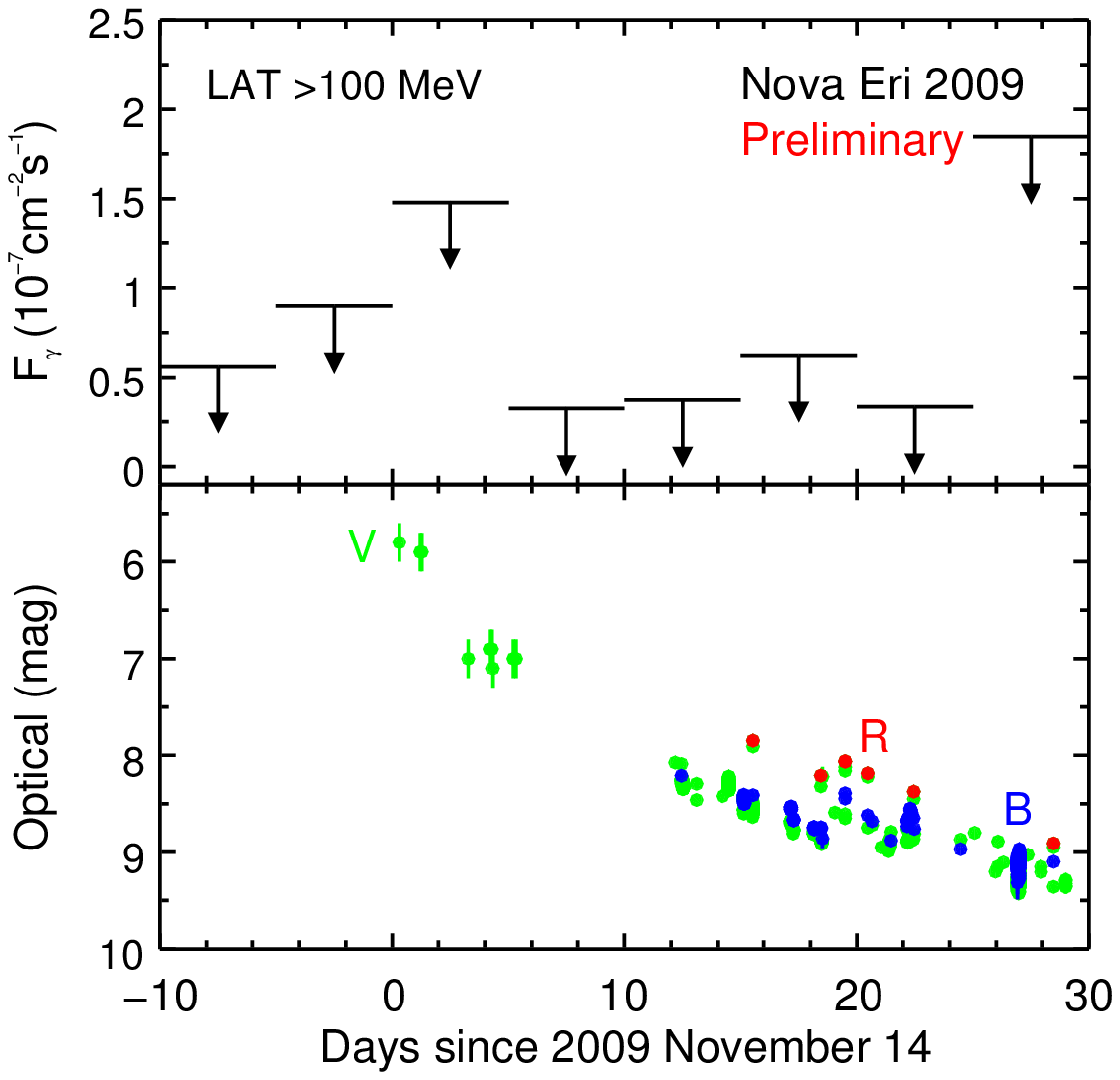}\includegraphics[width=88mm]{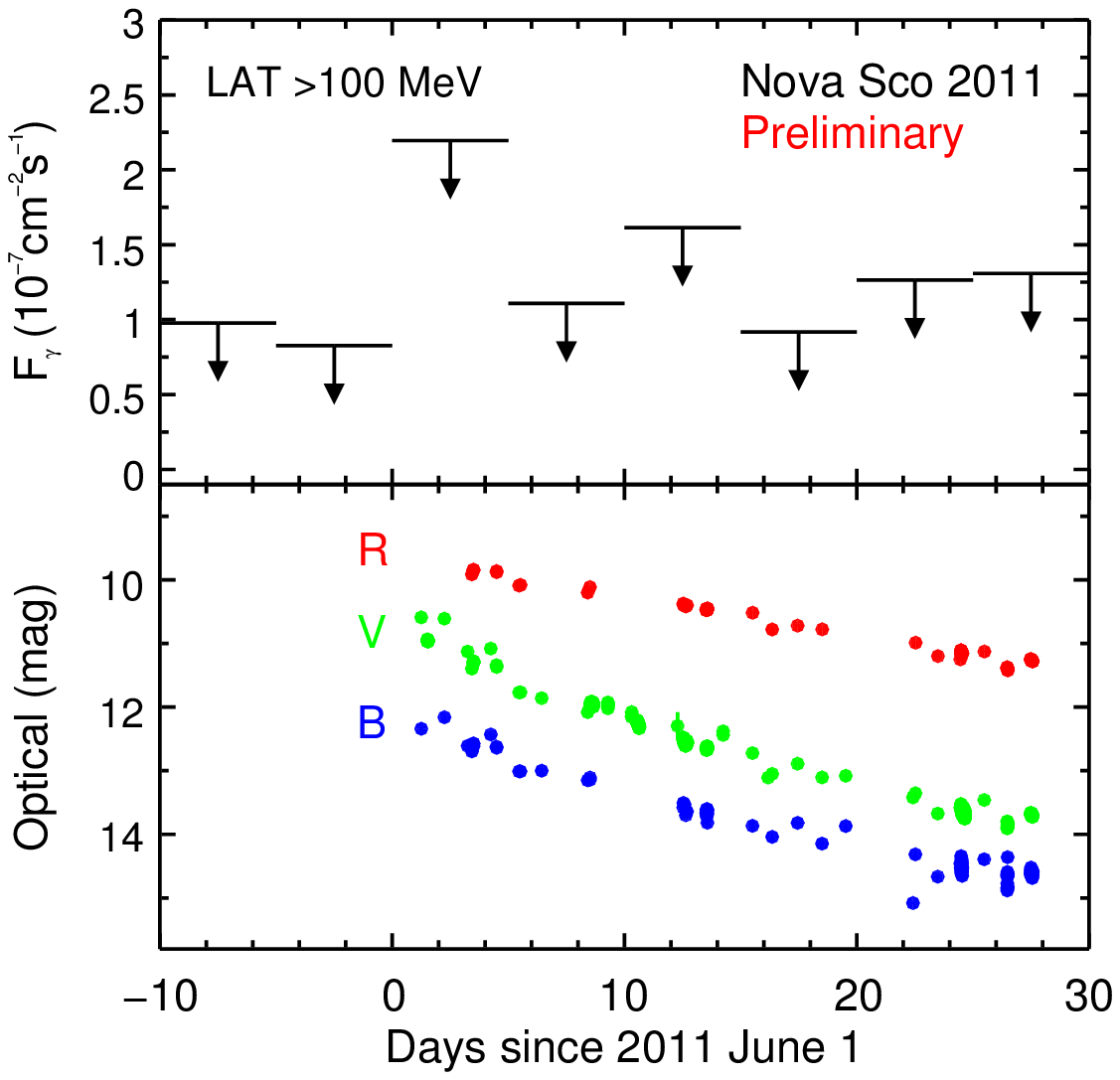}
\caption{\Fermi-LAT 5-day binned gamma-ray $2\sigma$ flux upper limits (top panels) and multi-color optical 
lightcurves from the AAVSO (bottom panels) for KT Eri 2009 (left) and V1312 Sco 
2011 (right).}\label{fig1}
\end{figure*}

\section{Interlude}

Since the V407 Cyg 2010 gamma-ray detection, the subsequent two years saw little 
observational progress on gamma-ray emission from novae because no other LAT 
detections were announced. This was in accord with expectations at the time, as 
the symbiotic-like recurrent novae are relatively rare, and as outlined in 
\cite{abd10}, the gamma-ray production mechanism appeared tied to the presence 
of a dense RG wind. This was enforced by the LAT search and resultant 
non-detections of other optically bright classical novae, binaries consisting of 
a white dwarf and main sequence star. Although they are much more common than 
the symbiotic-like recurrent novae (e.g., V407 Cyg; see Table~1), they may be 
fed by Roche lobe overflow rather than the massive winds from a RG companion. 
Fig.~\ref{fig1} shows two examples of such LAT non-detections of classical 
novae. In KT Eri 2009, the peak magnitude of 5.4 \cite{hou10} was brighter than 
observed in V407 Cyg 2010, while in another example, V1312 Sco 2011, the peak 
visual magnitude of 9.5 \cite{sea11} was typical of a rather faint nova 
detection. The LAT 5-day binned limits were $\simlt 10\times$ fainter than the 
1-day observed gamma-ray peak in V407 Cyg 2010. The distance to KT Eri is 6.5 
kpc \cite{rag09}, which is significantly larger than that commonly assumed for 
V407 Cyg ($\sim 2.7$ kpc; \cite{mun90}), and offers another limiting factor in 
explaining its non-detection with the LAT.

\begin{figure*}[t]
\centering
\includegraphics[width=130mm]{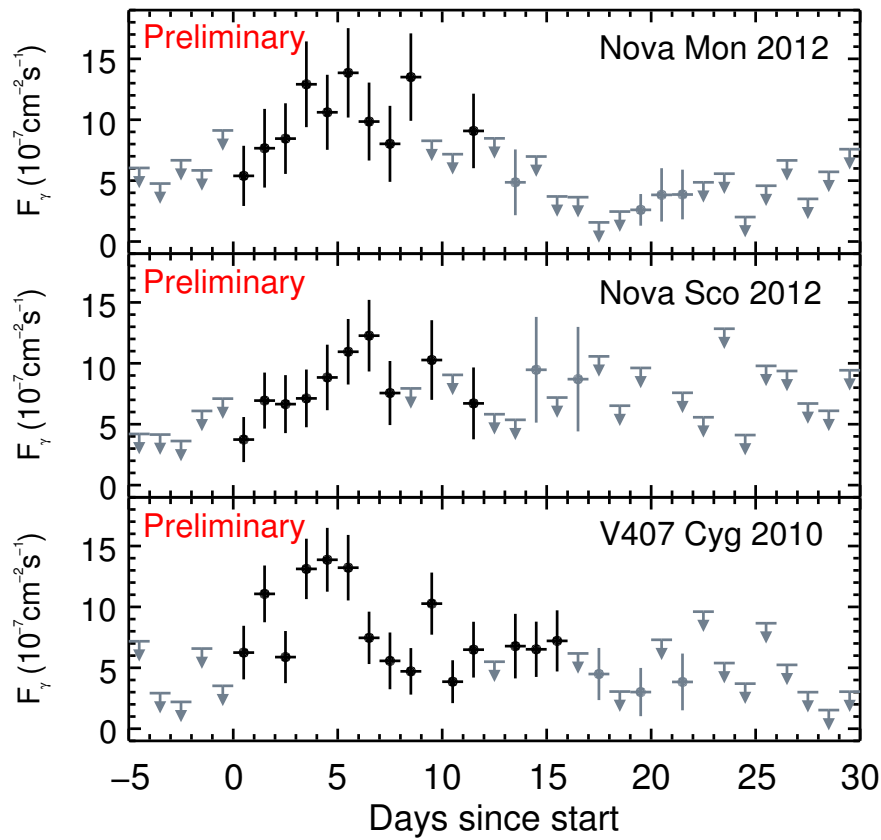}
\caption{\Fermi-LAT 1-day binned ($>100$ MeV) lightcurves of the three detected 
gamma-ray novae. The start dates indicated are from top to bottom, 2012 June 19, 
2012 June 15, and 2010 March 10. Detections with $\geq 3\sigma$ significances 
are the points with flux error bars shown in black, while data with $2-3\sigma$ 
significances are shown in gray; upper limits are shown for point with 
$<2\sigma$ significances.}
\label{fig2}
\end{figure*}

\section{Further Surprises in June 2012: Two New LAT Detected Novae}

In June 2012, the LAT detected two new novae, four days apart, and at the 
opposite ends of the sky. They followed different discovery paths, as outlined 
below. These were different from the V407 Cyg 2010 case in that these were 
classical novae, and offered new glimpses into the underlying high-energy 
particle acceleration physics responsible for gamma-ray emission in novae.

\subsection{Discovery of Nova Sco 2012}

Fermi J1750-3243 was discovered initially in LAT Routine Science Processing 
\cite{gla11} analysis in the field of an X-ray binary starting 2012 June 18 and 
confirmed with the Fermi All-sky Variability Analysis \cite{all12}. Subsequent 
analysis revealed the first gamma-ray daily detection was as early as June 16, with 
emission lasting for $\sim 2$ weeks \cite{che12a}. In the preliminary analysis 
(Fig.~\ref{fig2}), the gamma-ray ($>100$ MeV) peak of $\sim 10^{-6}$ \phflux\ 
was on June 21st while the initial detection was as early as the 15th; for 
further LAT details, see \cite{hil12}.

The Nova Sco 2012 optical discovery resulted from a microlensing survey of the 
Galactic bulge and was named MOA 2012 BLG-320. It exhibited a 6 mag brightening 
from June 1-3, with a gradual rise, peaking on June 19-21 \cite{wag12}, 
coincident with the gamma-ray emission peak/plateau emission seen in Fermi 
J1750-3243 (Fig.~\ref{fig2}). A 1.6 hr periodic modulation of 0.1 mag from May 
28-31 \cite{wag12} was observed although its nature is unclear. A large ejecta 
velocity is implied from infrared spectroscopy measuring a FWHM of 2200 km/s in 
the Pa$\beta$ line on June 17.879 \cite{raj12}. A faint radio source was also 
detected, and its steep spectrum is consistent with a synchrotron origin 
\cite{cho12a}. In this case, \Swift/XRT observations obtained starting June 22 
did not detect X-ray emission from the source \cite{pag12}, but as in the case 
of V407 Cyg (\S~1), analysis of the XRT image showed no other significant X-ray 
sources within the LAT error circle.

Nova Sco 2012 was classified as a classical nova \cite{wag12} which are 
typically accreting through Roche lobe overflow rather than from a RG wind as in 
the case of symbiotic-like recurrent nova. The recurrence time for CN are $\sim 
10^{4}$ years and the estimated rate is for $\sim 35$ such events per year in 
our Galaxy \cite{sha97}. (Contrast: only $\sim 10$ total recurrent nova are 
known.) This implied that CN could possibly be common as gamma-ray emitters, but 
still begs the question why this particular event and not others were detected 
by the LAT (cf., Fig~\ref{fig1}).

\begin{figure}
\includegraphics[width=80mm]{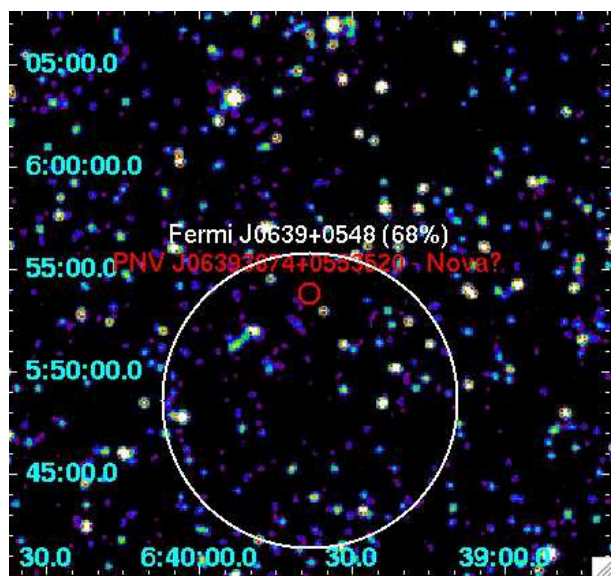}
\caption{Plot of the position of the reported possible nova, 
PNV~J06393874+0553520 (red circle; \cite{cbat}) on an archival DSS image with 
the Fermi J0639+0548 error circle \cite{che12b} overlaid. This initial 
`discovery' plot showed the positional coincidence between the optical and 
gamma-ray sources.}
\label{fig3}
\end{figure}

\subsection{Discovery of Nova Mon 2012, and its Fraternity}

\begin{figure}
\includegraphics[width=80mm]{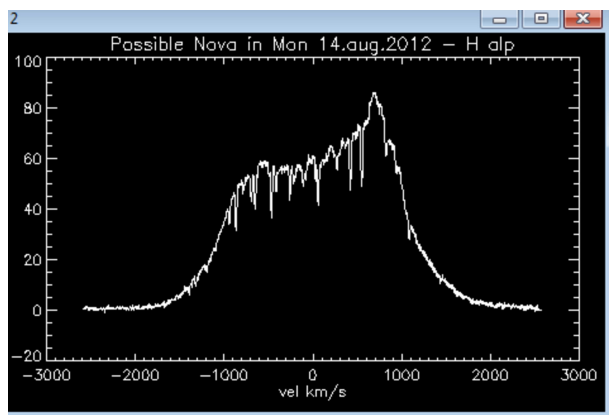}
\caption{H$\alpha$ profile of Nova Mon 2012 from the $R=15000$ spectrum obtained 
Aug 14, 2012 at the OHP (France) by S.~Charbonnel 
(image kindly provided by I.~De Gennaro Aquino). See \cite{aras2012} for more of 
the amateur spectra including the initial lower-resolution one from 
J.~Edlin.}
\label{fig4}
\end{figure}

Nova Mon 2012 was discovered initially by the LAT as an unidentified gamma-ray 
transient, Fermi J0639+0548, in late-June \cite{che12b}. VLA observations were 
triggered on the \Fermi\ detection, but the source was $\sim 20\deg$ from Sun at 
the time (in the Monoceros region of the Galaxy) and thus precluded optical and 
X-ray observations. Due to its similar long-duration in gamma-rays to the V407 
Cyg 2010 and Nova Sco 2012 cases (Fig.~\ref{fig2}), the possibility that the 
gamma-ray source could be a nova was noted internally\footnote{Aside: After the 
association of Fermi J1750--3243 with Nova Sco 2012 in late July, I began 
considering the possible nova origin for other unidentified gamma-ray transients 
(near and off the Galactic plane), and reasoned that \Swift-UVOT optical 
observations of Fermi~J0639+0548 when the source came out of its Sun constraint 
could test this idea. Instead, on August 14, after discussing some results with 
collaborator, S.N.~Shore, he noted in passing that he had {\it ``just activated 
a ToO at the NOT for the new nova in Mon, it's a pretty one but likely wasn't 
seen by Fermi (but you might check, discovered 9 Aug).''} Immediately when I 
heard the new nova was in Monoceros, I knew the Fermi J0639+0548 gamma-ray 
transient detected in June was it and made the scruffy picture shown in 
Fig.~\ref{fig3} to confirm, responding {\it ``I thought after V407 Cyg that LAT 
would be a great nova detector for such events so here it is -- the first 
'gamma-ray' discovery of a nova!''}} and indeed, the optical discovery of the 
possible nova was made on August 9 \cite{fuj12}. Coincidentally, there was a 
nova spectroscopy workshop held at the time at the Observatoire de Haute 
Provence (OHP) in France and the first amateur spectra were obtained by 
S.~Charbonnel (Durtal Observatory, France; Fig.~\ref{fig4}) and J.~Edlin (Idaho 
Falls, Idaho) on August 14, confirming the nova. The initial ispection by I.~De 
Gennaro Aquino of these optical spectra at the OHP and the high resolution 
Nordic Optical Telescope (NOT) spectrum from August 16 obtained by S.~N.~Shore 
indicated a similarity with the Oxygen-Neon (ONe) type classical nova V382 Vel 
1999 \cite{del02} approximately 50 days after outburst (Fig.~9.25 therein); see 
\cite{sho13} for details. This gave us confidence in associating the LAT 
transient from June with Nova Mon 2012 detected optically in August 
\cite{che12c}. This implied the optical nova was discovered in flat decline 
phase of its optical lightcurve, consistent with the initial optical magnitudes 
reported. We inferred peak 4.5 - 5.0 mag \cite{che12c} in June would have been 
naked eye and indeed it would have appeared as a new `visiting' star in the 
Monoceros constellation.

ONe type novae have the highest mass WDs with massive and fast ejecta. The 
distance to Nova Mon 2012 is about 3.6 kpc \cite{sho13} and a detected 
periodicity of 7.1 hr \cite{osb13,wag13} is suggestive of the binary orbital 
period. As mentioned above, early radio detections (before the optical 
discovery) were possible due to the VLA observations triggered on the LAT 
discovery, starting $\sim 9$ days after the gamma-ray detection \cite{cho12b}. 
The radio emission exhibited a flat spectrum initially \cite{cho12b} and evolved 
with a sharply rising ($\nu^{1.7}$, \cite{fuh12}) spectrum that indicates 
optically thick Bremsstrahlung emission in the ejecta. The radio structure was 
resolved into a double source \cite{obr12} indicative of a bi-polar structure, 
confirmed by the optical line profiles \cite{sho13}. After the optical 
discovery, and \Fermi\ association, an X-ray source was discovered with \Swift\ 
\cite{nel12a} and has since gone through its supersoft brightening \cite{nel12b} 
and decline \cite{pag13}.

\section{Closing Thoughts}

The LAT properties for all three gamma-ray detected novae are remarkably 
similar, with soft spectra and emission detected up to $\sim$few GeV. The 
gamma-ray lightcurves (Fig.~\ref{fig2}) all similarly show fluxes rising over 
several days with peaks of $\approx 10^{-6}$ ph cm$^{-2}$ s$^{-1}$ ($E > 100$ 
MeV) over consecutive days and overall, are comparably long-duration ($\sim 2$ 
weeks at $> 3\sigma$, up to $\sim 3$ weeks at $> 2\sigma$). At face value, this 
suggests a common gamma-ray emission mechanism. However, their underlying binary 
systems are dramatically different, so the gamma-ray emission mechanisms are not 
necessarily the same. Fermi acceleration in the nova shell and interaction with 
the massive RG wind was determined to play an important role in V407 Cyg 
\cite{abd10} and the necessary conditions appear to be a massive WD and fast and 
massive ejecta. In comparison with V407 Cyg, the lack of dense environment in 
the two June 2012 gamma-ray novae make it difficult to explain the origin of the 
putative shocks that accelerate particles to high energies. The origin of the 
gamma-ray emission in CN are thus unclear -- whether hadronic or leptonic, and 
the particle acceleration sites. The resemblance between Nova Mon 2012 and other 
historically bright ONe CN (V1974 Cyg 1992, V382 Vel 1999, and Nova LMC 2000, 
together dubbed ``The Fabulous Four''; \cite{sho13}) is striking, and provides 
some early clues in these systems. Future LAT detections (and limits as well) of 
gamma-rays from novae will help in our understanding and improve our knowledge 
of the physics of shocks in general.

The initial association of the Fermi~J2102+4542 source with nova V407 Cyg 2010 
met with some skepticism, simply because nova were not widely considered as 
potential HE gamma-ray sources. Galactic transients in general have been rare to 
detect with the LAT and particular regions in the Galaxy are difficult to study 
due to the bright diffuse emission. The main question after the V407 Cyg nova 
discovery was whether it was the only nova \Fermi\ had/will see, and what other 
new transient phenomena will be detected. With the two new CN detections in June 
2012, novae are now established as a new gamma-ray source type after 4+ years of 
\Fermi\ surveying. Continued LAT searches thus have immense discovery potential, 
including the possible identification of newer, even more rare types of cosmic 
accelerators, foreseen or unforeseen.

\begin{acknowledgments}
The \textit{Fermi} LAT Collaboration acknowledges support from a number of 
agencies and institutes for both development and the operation of the LAT as 
well as scientific data analysis. These include NASA and DOE in the United 
States, CEA/Irfu and IN2P3/CNRS in France, ASI and INFN in Italy, MEXT, KEK, and 
JAXA in Japan, and the K.~A.~Wallenberg Foundation, the Swedish Research Council 
and the National Space Board in Sweden. Additional support from INAF in Italy 
and CNES in France for science analysis during the operations phase is also 
gratefully acknowledged.

We acknowledge with thanks the variable star observations from the AAVSO 
International Database contributed by observers worldwide and used in this 
research, and the dedicated observers of the Astronomical Ring for Access to 
Spectroscopy (ARAS; \cite{aras2012}) for their critical data on Nova Mon 2012 
and V407 Cyg 2010.

This research was supported at NRL by a Karles' Fellowship, NASA DPR S-15633-Y, 
and by the \textit{Fermi} guest investigator program. The author is grateful to 
Steve Shore for his continued guidance and collaboration in novae work,
and to Davide Donato, Liz Hays, Pierre Jean, and Tyrel Johnson, and Steve Shore 
for comments on the text.
\end{acknowledgments}


\end{document}